\documentclass[10pt,letterpaper]{article}
\usepackage[top=0.55in,left=1.0in]{geometry}

\usepackage[utf8]{inputenc}

\usepackage{cite}

\usepackage{bm}
\usepackage{amsmath}
\usepackage[right]{lineno}

\usepackage{microtype}
\DisableLigatures[f]{encoding = *, family = * }

\usepackage{changepage}

\usepackage[aboveskip=1pt,labelfont=bf,labelsep=period,singlelinecheck=off]{caption}

\makeatletter
\renewcommand{\@biblabel}[1]{\quad#1.}
\makeatother

\usepackage{lastpage,fancyhdr,graphicx}
\usepackage{epstopdf}
\fancyheadoffset[L]{0.25in}
\fancyfootoffset[L]{0.25in}

\usepackage{color}

\definecolor{Gray}{gray}{.25}

\usepackage{graphicx}

\usepackage{wrapfig}
\usepackage[pscoord]{eso-pic}
\usepackage[fulladjust]{marginnote}
\reversemarginpar

\begin{document}
\vspace*{0.05in}

\begin{flushleft}
{\large
\bf{{A Numerical Fitting Routine for Frequency-domain Thermoreflectance Measurements of Nanoscale Material Systems having Arbitrary Geometries}}
}
\newline
\\
Ronald J. Warzoha\textsuperscript{1,*},
Adam A. Wilson\textsuperscript{2},
Brian F. Donovan\textsuperscript{3},
Andrew N. Smith\textsuperscript{1},
Nicholas T. Vu\textsuperscript{1},
Trent Perry\textsuperscript{1},
Longnan Li\textsuperscript{4},
Nenad Miljkovic\textsuperscript{4}, and
Elizabeth Getto\textsuperscript{1}
\\
\bigskip
\textsuperscript{\bf{1}}Department of Mechanical Engineering, United States Naval Academy, Annapolis, MD 21402, USA
\\
\textsuperscript{\bf{2}}Research Scientist, CCDC Army Research Laboratory Adelphi, MD 20783, USA
\\
\textsuperscript{\bf{3}}Department of Physics, United States Naval Academy, Annapolis, MD 21402, USA
\\
\textsuperscript{\bf{4}}Department of Mechanical Science and Engineering, University of Illinois, Champaign, IL 61821, USA
\\
\bigskip
* Corresponding Author, warzoha@usna.edu

\end{flushleft}

\section*{Abstract}
In this work, we develop a numerical fitting routine to extract multiple thermal parameters using frequency-domain thermoreflectance (FDTR) for materials having non-standard, non-semi-infinite geometries. The numerical fitting routine is predicated on either a 2-D or 3-D finite element analysis that permits the inclusion of non semi-infinite boundary conditions, which can not be considered in the analytical solution to the heat diffusion equation in the frequency domain. We validate the fitting routine by comparing it to the analytical solution to the heat diffusion equation used within the wider literature for FDTR and known values of thermal conductivity for semi-infinite substrates (SiO$_2$, Al$_2$O$_3$ and Si). We then demonstrate its capacity to extract the thermal properties of Si when etched into micropillars that have radii on the order of the pump beam. Experimental measurements of Si micropillars with circular cross-sections are provided and fit using the numerical fitting routine established as part of this work. Likewise, we show that the analytical solution is unsuitable for the extraction of thermal properties when the geometry deviates significantly from the standard semi-infinite case. This work is critical for measuring the thermal properties of materials having arbitrary geometries, including ultra-drawn glass fibers and laser gain media. 


\section*{Introduction}
Measurements of thermal transport properties in nanoscale thin-films are conventionally made using optical pump-probe thermoreflectance techniques, principally due to the non-contact nature with which they are able to interrogate nanoscale thermal transport characteristics for nearly any material type \cite{jiang2018tutorial}. In contrast to Raman spectroscopy \cite{yan2014thermal} and 3-$\omega$ \cite{tong2006reexamining} techniques, thermoreflectance-based measurements can separate the impacts of thermal boundary conductance (G) across interfaces and thermal conductivity ($\kappa$) within individual material layers \cite{kang2008two} and have overwhelmingly served as the thermal characterization technique of choice for nanoscale material systems over the course of the last decade \cite{jiang2018tutorial,feser2014pump,ge2006thermal,lyeo2006thermal,losego2012effects}. However, current models used to extract the thermal properties of nanoscale materials limit the geometries that can be interrogated to those which are (1) semi-infinite in the radial direction and (2) have finite or semi-infinite thickness transverse to the direction of the applied heat source \cite{cahill2004analysis}. In this work, we establish a finite element-based numerical fitting routine in order to extend the utility of thermoreflectance techniques for use with any planar geometry having finite dimensions.

The two thermoreflectance systems most used over the course of the last decade include time-domain thermoreflectance (TDTR) \cite{zhu2010ultrafast} and frequency-domain thermoreflectance (FDTR) \cite{schmidt2009frequency}. Both techniques rely on two separate laser beams to (1) heat the sample surface and (2) probe the reflectivity (i.e. temperature) on the sample surface. The beams that heat and probe the surface are referred to as the ``pump'' and ``probe'' beams, repsectively. Typically, a metal transducer (50-150 nm of Au or Al) is deposited on the sample surface in order to convert the optical energy of the pump beam to thermal energy prior to the sample surface {\it and} due to a well-established wavelength-dependent relationship between the reflectivity of the transducer and its surface temperature (often referred to as the coefficient of thermoreflectance) \cite{favaloro2015characterization,wilson2012thermoreflectance}. The pump beam is modulated at a single frequency (TDTR) or across a range of frequencies (FDTR) such that we can use a lock-in amplifier to detect small changes in the reflectance as heat penetrates into the sample. In TDTR, a pulsed laser source is used and split into two different pump and probe paths. A delay stage is used to physically delay the arrival of the probe beam relative to the arrival of the pump beam to monitor changes in reflectance (i.e. temperature) at the sample surface over time. On the other hand, FDTR utilizes two separate continuous wave (CW) lasers to establish pump and probe beams, where the pump beam is modulated over a range of frequencies. Modulating frequency allows for corresponding changes to the penetration depth of the heat deposited by the pump beam and thus establishes sensitivity to multiple thermal properties and/or the thermal properties of several underlying layers of material in a multi-layer stack \cite{cahill2004analysis}.

Recent advances in thermoreflectance-based techniques include extensions to a steady-state system to gain sensitivity to so-called ``buried interfaces'' \cite{braun2019steady}, the use of a magneto-optical kerr effect (MOKE) to gain sensitivity to interfaces having large thermal conductance \cite{kimling2017thermal} and phonon-magnon coupling effects \cite{chen2017effects} and the development of a transient thermo-transmission technique to measure the thermal boundary conductance of nanoparticles suspended in transparent media \cite{szwejkowski2017molecular}. Recently, transient grating spectroscopy has also been used to probe non-diffusive thermal regimes in films that are suspended over micron-sized regions \cite{johnson2013direct,kim2017elastic,minnich2012determining}. However, conventional measurements are still limited to geometries that are semi-infinite, which still limits the utility of the technique to material systems that can be fabricated in such a way. For instance, these characterization techniques are blind to the thermal properties of material systems whose thermal properties are expected to {\it change} with geometry, such as ultra-drawn glass fibers (i.e. fiber-optic components), strained polymers and laser gain media. In this work, we integrate a finite element-based numerical simulation (constructed in COMSOL Multiphysics v. 5.5) into the conventional fitting routine for frequency-domain thermoreflectance measurements such that the thermal properties of nanoscale and microscale material systems having non-standard geometries can be accurately extracted. We take measurements on standard substrates having well-known thermal properties (SiO$_2$, Al$_2$O$_3$ and Si) to validate the numerical model against values obtained analytically and to those that exist within the wider literature. We then demonstrate the difference in the numerical solution to the phase lag (i.e. temperature response) at the sample surface due to changes in the radial geometry of Si in the form of Si micropillars with varying height. Finally, we obtain the thermal conductivity of Si when in micropillar form using the numerical fitting routine and an experimental measurement made using FDTR.

\section*{Experiment}

\begin{wrapfigure}[22]{r}{90mm}
    \centering
    \includegraphics[scale=0.65]{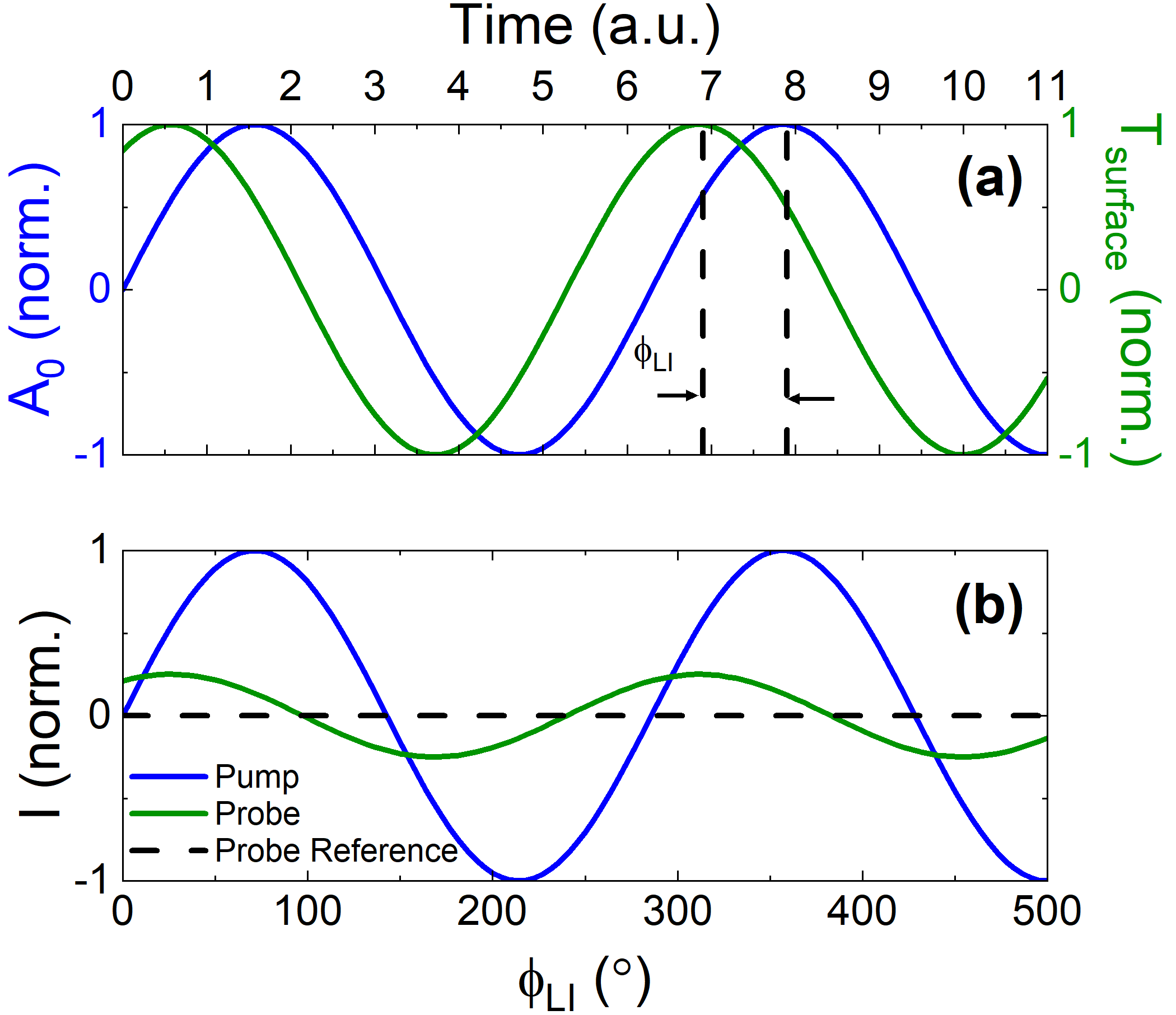}
    \caption{\small{\color{Gray}({\it a}) Time-domain representation of modulated heat source (blue) and temperature response (green), where the peak-to-peak difference is the phase lag as measured by the lock-in amplifier ($\phi_{LI}$) and ({\it b}) Phase lag vs. measured beam intensity. We note that the temperature difference between the probe beam and the probe reference (i.e. the temperature rise/fall at the transducer surface in response to the modulated heating event) is very small, requiring the use of a lock-in amplifier).}}
    \label{phiLa}
\end{wrapfigure}

In this work, FDTR is used to characterize the thermal properties of bulk substrates and Si micropillars. Past works provide a detailed description of the FDTR system used in this effort \cite{warzoha2019nanoscale,donovan2019elimination}. Briefly, FDTR measures the phase lag at the sample surface relative to the imposed phase applied by the modulated heating event. One useful analogy (despite occurring in the time-domain) is the lag in the temperature response of water relative to the temperature of a stove-top's heater. Because our experiment applies a sinusoidal modulation to the heating event, we can observe the temperature response at the sample surface by measuring the phase lag at the sample surface. A representative schematic of the mechanism used to track temperature on the sample surface is provided in Fig. \ref{phiLa}.

\begin{figure*}[t]
    \centering
    \includegraphics[scale=0.6]{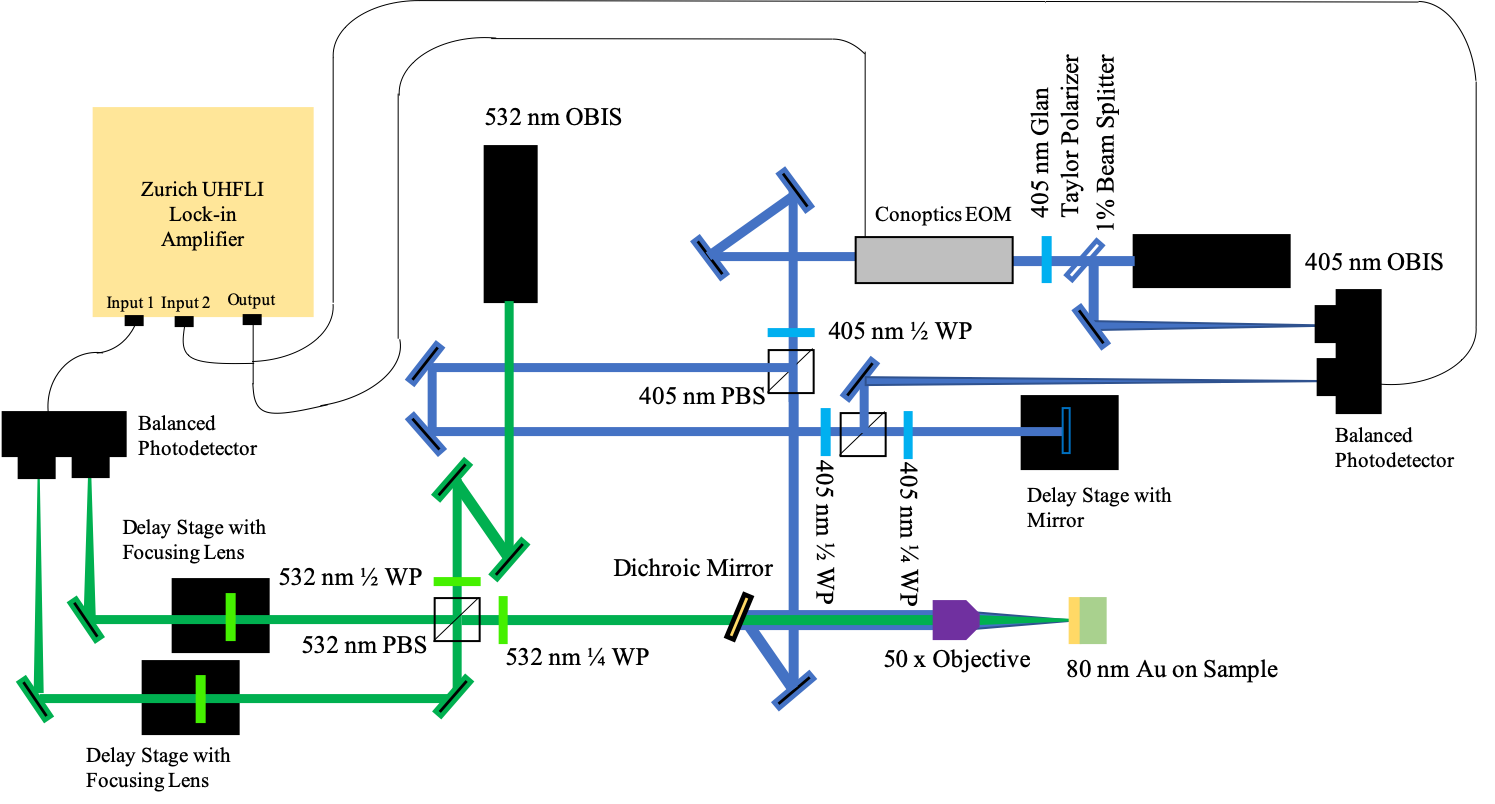}
    \caption{\small{\color{Gray}Schematic of frequency-domain thermoreflectance system built at USNA. Note the following acronyms: PBS (polarizing beam splitter), WP (waveplate) and EOM (electro-optic modulator).}} 
    \label{schem}
\end{figure*}

In Fig. \ref{phiLa} (a), the blue curve represents the power of the pump laser, which is modulated between an upper and lower power (1 and -1, respectively).  The corresponding temperature response of the surface is shown in green, which lags behind the applied temperature and fluctuates between and upper and lower temperature (again 1 and -1, respectively).  The difference in the phase between the pump and probe is the phase lag measured by the lock-in amplifier, $\phi_{LI}$, and this provides information on the thermal properties of the material when used in tandem with a well-established multi-layer analytical model \cite{schmidt2009frequency}.

Figure \ref{phiLIa} (b) displays the intensity of the pump beam as well as the intensity of the reflected probe.  The intensity and apparent modulation of the probe beam comes from the temperature response of the material induced by the pump.  The intensity of the reflected probe depends on the surface reflectivity, which is proportional to the change in temperature.  The probe reference that did not interact with the surface---and therefore remains unmodulated---is shown by the dotted black line.  By subtracting the probe signal from the reference and taking into account the coefficient of thermoreflectance, the temperature response of the sample can be determined.

A schematic of the FDTR system used in this work is shown in Fig. \ref{schem}. The system contains two separate continuous-wave lasers that act as the ``pump'' (405 nm Coherent OBIS CW laser) and ``probe'' (532 nm Coherent OBIS CW laser) respectively. We note that at these wavelengths, the transducer absorbs a significant amount of the pump beam and we are highly sensitive to changes in the reflectivity of the sample surface by the probe beam.

The pump beam is first split into two separate paths: (1) through the electro-optic modulator (EOM, Conoptics Model 350-160) and (2) into the balanced photodetector (Thorlabs, PDB450A-AC) that tracks the reference phase via a 1$\%$ beam splitter. The small amount of pump beam that is immediately directed into the photodetector is used as a reference to subtract any coherent noise within the laser. The portion of the pump beam that is routed through the EOM is modulated using the built-in waveform generator in our lock-in amplifier (Zurich UHFLI). After exiting the EOM, the pump beam is again spit into separate ``primary'' (downward direction) and ``reference'' (leftward direction) paths using a polarizing beam splitter. The primary path is steered into an objective lens (Mitutoyo 50x) using a dichroic mirror and focused onto the sample surface. The reference path is used to track the applied phase of the pump beam at the sample surface. In order to match the phase at the sample surface, we modulate at the highest frequency used in our measurement ($\omega_{max}$ = 20 MHz) and match the numerical phase of the pump beam that leaks into the primary balanced photodetector (shown on the left of the image) while blocking the probe beam. This allows us to measure the phase response at the sample surface (measured with the probe beam) and the imposed phase at the sample surface (applied via the pump beam) {\it simultaneously}, which is unique to our implementation of FDTR. By subtracting the two signals, we obtain $\phi_{LI}$ as described previously. 

Provided with a measurement of $\phi_{LI}$, one can obtain the underlying thermal properties across an interface (i.e. the thermal boundary conductance, G) or within an individual layer of a multilayer stack (e.g. the thermal conductivity, $\kappa$). The entire formulation of the analytical expression used to extract thermal properties is described in detail elsewhere \cite{schmidt2009frequency}. However, it is useful to describe several features of the analytical solution to the frequency-domain version of the heat diffusion equation in order to demonstrate its geometric limitations. The analytical solution is constructed within the framework of a multi-layer material stack whose substrate is semi-infinite in the radial direction, and most often semi-infinite in the through-thickness direction. A schematic of the general multilayer material stack used in the development of the analytical solution is provided in the figure below.

\begin{figure} [h!]
    \centering
    \includegraphics[scale=0.82]{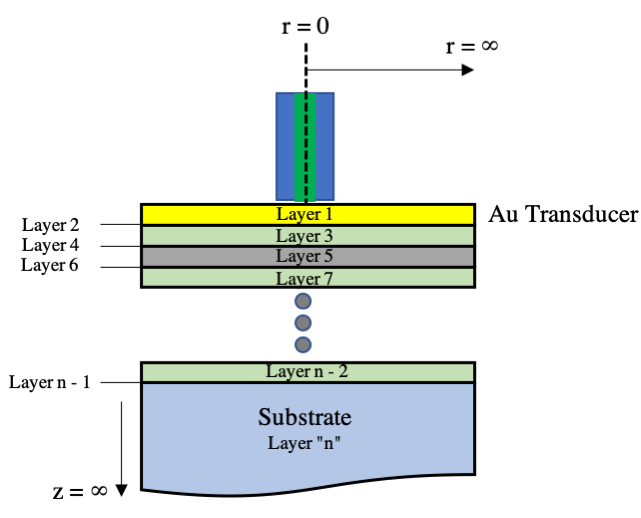}
    \caption{\small{\color{Gray}Arbitrary multi-layer material stack with semi-infinite boundary conditions in r- and z-directions. Pump (blue) and probe (green) beams are depicted above the Au transducer.}}
    \label{fig:my_label}
\end{figure}

In the time domain, the equation governing heat diffusion is expressed as,

\begin{equation}
    C_v \frac{\partial \theta}{\partial t} = \frac{\kappa_{r}}{r} \frac{\partial}{\partial r} \bigg(r \frac{\partial \theta}{\partial r}\bigg) + \kappa_{z} \frac{\partial^{2} \theta}{\partial z^{2}}
\end{equation}

\noindent The above expression accounts for 2-D heat flow in the radial (r) and through-plane (z) directions. A Fourier transform is applied to obtain the frequency-dependent heat diffusion equation, written as,

\begin{equation}
    \kappa_{z} \frac{\partial^{2} \theta (\omega,k,z)}{\partial z^{2}} = (\kappa_{r} k^{2} + C_{v} i \omega)\theta(\omega,k,z)
\end{equation}

\noindent where $q$ is defined for a layer of material $n$ and thickness $d$ as   

\begin{equation}
    q^2= \frac{\kappa_r k^2 + C_v i \omega}{\kappa_z}
\end{equation}

\noindent We can relate the temperature to the heat flux at the top surface (subscript $t$) of a slab made of a certain material in the frequency domain with the bottom surface (subscript $b$) using,

\begin{equation}
\left[\begin{matrix}
\theta_{n,b}\\
f_{n,b}
\end{matrix}\right]
= 
\left[\begin{matrix}
cosh(qd) & -\frac{1}{\kappa_z q}sinh(q d) \\
-\kappa_z *q*sinh(q d) & cosh( q d)
\end{matrix}\right]
\left[\begin{matrix}
\theta_{n,t}\\
f_{n,t}
\end{matrix}\right]
\end{equation}

\noindent The temperature and heat flux between the bottom surface of material $n$ are connected to the top of material $n+1$ via  

\begin{equation}
\left[\begin{matrix}
\theta_{n+1,t}\\
f_{n+1,t}
\end{matrix}\right]
=
\left[\begin{matrix}
1 & -G^{-1}\\
0 & 1
\end{matrix}\right]
\left[\begin{matrix}
\theta_{n,b}\\
f_{n,b}
\end{matrix}\right]
\end{equation}

\noindent where $G$ is the thermal boundary conductance between the two layers.  The heat flux boundary condition of the top, $f_t$, can be found with  

\begin{equation}
    f_t = \frac{A_0}{2\pi} exp\left(-\frac{k^2 w_o^2}{8}\right)
\end{equation}

\noindent which is the Hankel transform of a Gaussian spot with a power of $A_0$ and a $1/e^2$ radius of $w_0$. If there are multiple layers, the solution can be found with

\begin{equation}
\left[\begin{matrix}
\theta_{b}\\
f_{b}
\end{matrix}\right]
=\boldsymbol{M_n}\boldsymbol{M_{n-1}}...\boldsymbol{M_2}\boldsymbol{M_1}
=
\left[\begin{matrix}
A & B\\
C & D
\end{matrix}\right]
\left[\begin{matrix}
\theta_{t}\\
f_{t}
\end{matrix}\right]
\end{equation}

\noindent where $\boldsymbol{M_n}$ is the matrix of the bottom layer. If the bottom layer is treated as adiabatic or semi-infinite, the surface temperature can be found using:

\begin{equation}
    \theta_t = -\frac{D}{C}f_t
\end{equation}

\noindent The final frequency response, $H(\omega)$, is found by taking the inverse Hankel transform of Eqn. 2 and weighting it with a Gaussian spot with a $1/e^2$ radius of $w_1$:  

\begin{equation}
    H(\omega) = \frac{A_0}{2\pi} \int_{0}^{\infty} k \left( -\frac{D}{C} \right) exp\left(-\frac{k^2 ({w_0}^2 + {w_1}^2}{8}\right) dk
\end{equation}

\begin{wrapfigure}[23]{r}{75mm}
    \centering
    \includegraphics[scale=0.8]{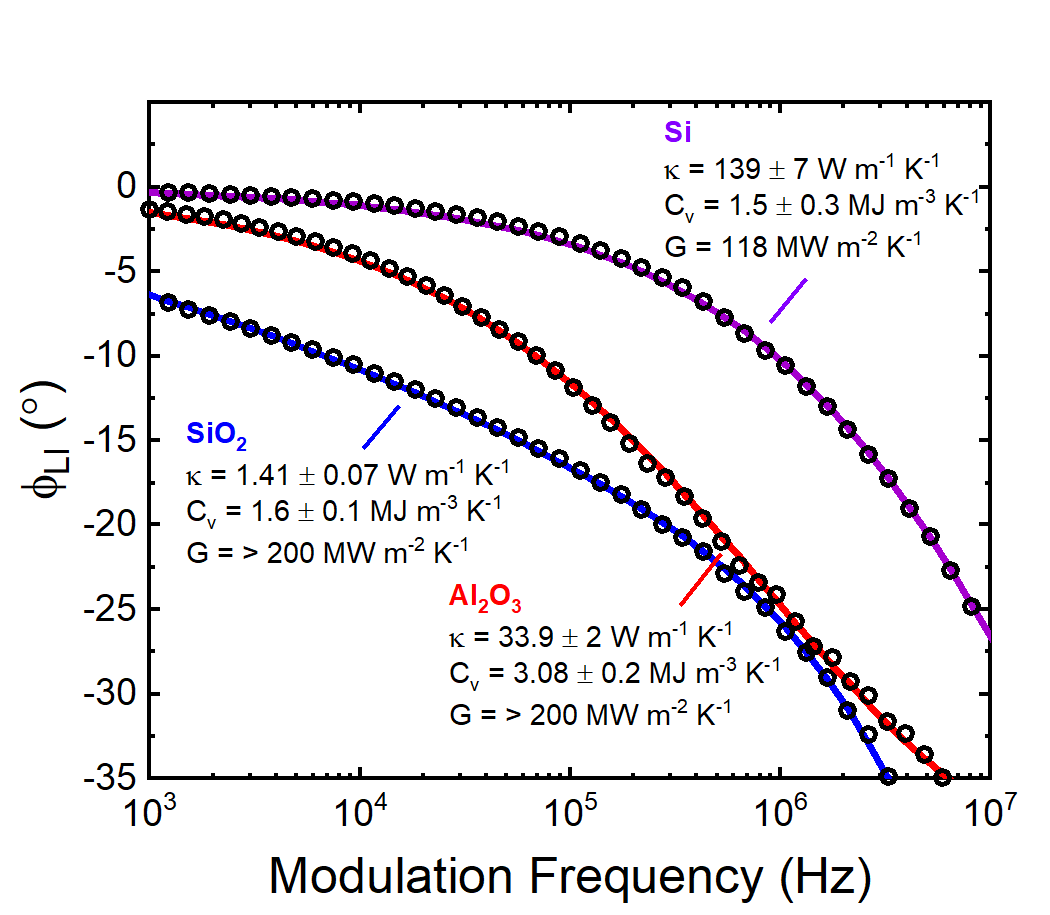}
    \caption{\small{\color{Gray}Measured data (open black circles) and corresponding analytical fits for SiO$_2$ (solid blue line), Al$_2$O$_3$ (solid red line) and Si (solid purple line) substrates. Pump and probe diameters are 5.7 $\mu$m and 3.7 $\mu$m, respectively.}}
    \label{substrates}
\end{wrapfigure}

\noindent The thermal model for $H(\omega)$ is then fitted to the lock-in phase data.  By changing the parameters of the thermal model to fit the lock-in data, the thermal properties can be determined.  The lock-in phase data measured is given by

\begin{equation}
    \phi_{LI}=tan^{-1} \frac{\Im (H(\omega))}{\Re (H(\omega))} + \phi_{ext}
    \label{phiLI}
\end{equation}

\noindent where $\Im (H(\omega))$ is the out-of-phase signal, $\Re (H(\omega))$ is the reference signal, and $\phi_{ext}$ is the external phase shift caused by other aspects not caused by changes in reflectively, such as the optical path length, driving electronics, and photodetectors. Thermal properties are extracted by fitting Eqn. \ref{phiLI} to measurements of the phase lag via FDTR for SiO$_2$, Al$_2$O$_3$ and Si, shown in Fig. \ref{substrates}, below.

\noindent The thermal properties obtained above are consistent with those widely reported in the scientific literature \cite{braun2019steady}. The thermal boundary conductance for all three samples is high relative to other values reported in the literature \cite{braun2019steady}; however, we utilize an $\sim$ 5 nm Ti layer between the Au transducer and the bulk substrates, which improves adhesion and therefore enhances thermal transport across the interface. 

\begin{wrapfigure}[31]{r}{75mm}
    \centering
    \includegraphics[scale=0.8]{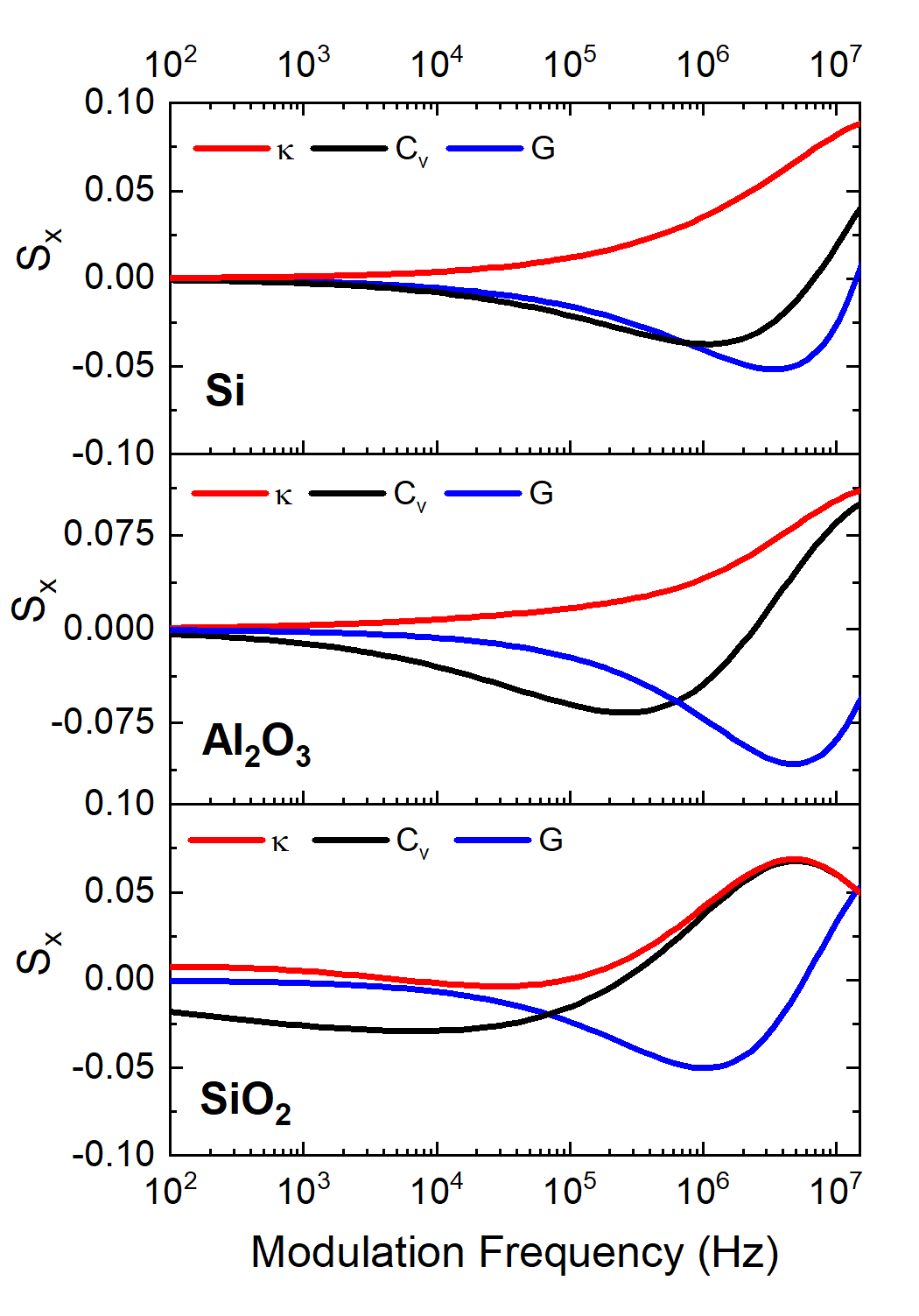}
    \caption{\small{\color{Gray}Sensitivity (S) to the extracted thermal properties (where each property is represented by the subscript ``x'') for each substrate shown in Fig. \ref{substrates}.}}
    \label{sens1}
\end{wrapfigure}

To determine whether a particular thermal property ($\kappa$, C$_v$ and G) can be extracted across the range of modulation frequencies applied on the sample surface, we can determine the sensitivity of each parameter to small perturbations in the measured phase lag.

The phase sensitivity to a particular thermal property, $x$, at a given frequency can be found with

\begin{equation}
    S(\omega)=\frac{\partial \phi (\omega)}{\partial \ln x}
\end{equation}

\noindent The sensitivity to the thermal properties shown in Fig. \ref{substrates} is provided in Fig. \ref{sens1}.

A two-dimensional (2D) finite element model is developed in order to extract thermal properties from geometries with non semi-infinite boundary conditions. The most palatable system to demonstrate the utility of the finite element model is one in which the geometry is confined in the radial direction and one that has been widely fabricated in laboratory environments. Consequently, we choose to develop the model based on Si micropillar arrays. The arrays we interrogate have geometries that vary in both the radial and through-plane directions on the order of single-digit $\mu$m to 10's of $\mu$m. As these length remain larger than the mean free path of phonons in Si \cite{wang2011thermal}, the geometric confinement should {\it not} result in any change in $\kappa_{Si}$. However, it is likely that the {\it phase lag ($\phi_{LI}$) does change} due to the confinement of heat (recall that the phase lag represents the response in the temperature on the top of the transducer relative to the modulated signal of the applied heating event).

\begin{wrapfigure}[24]{r}{90mm}
    \centering
    \includegraphics{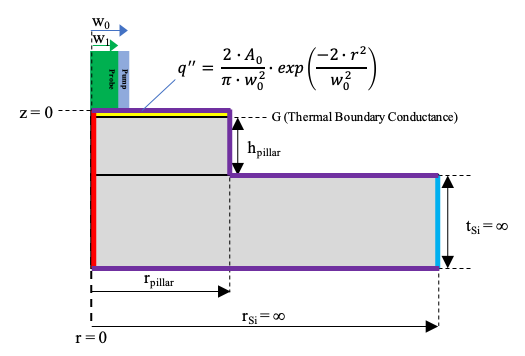}
    \caption{\small{\color{Gray}Schematic of numerical model built in COMSOL multiphysics. h$_{pillar}$ represents the pillar height, t$_{Si}$ is the thickness of the Si substrate (approximated as semi-infinite, but modeled as 500 $\mu$m in the computational domain), r$_{pillar}$ is the radius of the micropillar, r$_{Si}$ is the radius of the Si substrate (approximated as semi-infinite, but modeled as 300 $\mu$m) and w$_0$ and w$_1$ are the pump and probe radii, in $\mu$m. Purple boundaries represent those boundaries that are insulated, blue boundaries represent boundaries which are held at a constant initial temperature, T$_0$ = 300 K and red boundaries are symmetric.}}
    \label{numschem}
\end{wrapfigure}

The finite element model is created in COMSOL Multiphysics v. 5.5. We note here that COMSOL is used with its LiveLink module in order to communicate with Matlab. The Si micropillar is modeled using the computational domains shown in Fig. \ref{numschem}. We note that we incorporate the Si substrate {\it and} the Si micropillar within our model. An 80 nm Au transducer is constructed above the pillar and a finite value for thermal boundary conductance is applied at the Au/Si micropillar interface (in our model this remains a free parameter). The Si micropillar and the Si substrate are ``continuous'' in the sense that there is no applied thermal boundary conductance at the interface between domains. This is physically appropriate given the nature of the fabrication process; the Si micropillars themselves are etched and are never physically separated/reattached during the process.

The numerical model itself is meshed using a graded grid in each independent sub-domain shown in Fig. \ref{numschem}, including the Au transducer layer, the Si micropillar, the region immediately below the Si micropillar (i.e. from r = 0 to r = r$_{pillar}$ in the Si substrate) and the remainder of the Si substrate. An image of the meshed sub-domains is provided in Fig. \ref{mesh}. Note that a mesh independence study was completed to ensure that the solution was independent of the number of nodal points in each sub-domain.

\section{Numerical Methods}

As shown, the mesh size (i.e. the size of each mesh element) is increased downward (z $\rightarrow$ $\infty$) and to the right (r $\rightarrow$ $\infty$) in sub-domains 2, 3 and 4. However, the mesh size {\it decreases} in the downward direction (z $\rightarrow$ $\infty$). This is done in an effort to capture the relevant thermal transport physics at the transducer/pillar interface (i.e. the thermal boundary conductance, G). This is particularly important when (1) we are sensitive to G and (2) we need to extract $\kappa$ independent of G (as we do here). Prior to using the numerical simulation described here to fit any data, we use it to fit to the data in Fig. \ref{substrates}.

\subsection{Validation of Numerical Models}

\begin{wrapfigure}[24]{r}{90mm}
    \centering
    \includegraphics[scale=0.55]{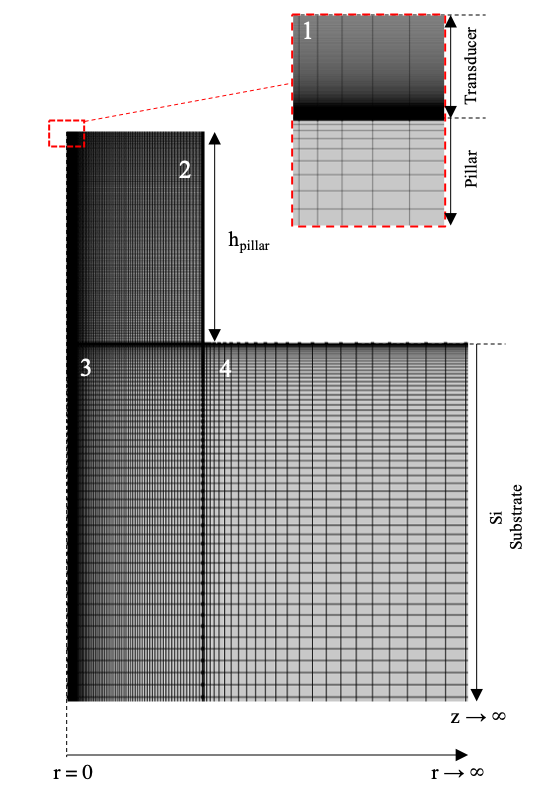}
    \caption{\small{\color{Gray}Mesh for Si micropillar on Si substrate with Au transducer. Entire domain contains 64,000 elements split equally among four distinct sub-domains (numbered 1-4).}}
    \label{mesh}
\end{wrapfigure}

The numerical model described in the previous section is validated by fitting the measured data in Fig. \ref{substrates} and comparing it to both the resulting analytically determined thermal properties and those available within the wider literature. The only changes to the models shown in Figs. \ref{schem} and \ref{mesh} include: (1) removal of the pillar domain and (2) the transducer is split into two separate domains above the Si substrate (one from r = 0 to r = 2$\cdot$(w$_0$ + w$_1$) and the other from r = 2$\cdot$(w$_0$ + w$_1$) to r $\rightarrow$ $\infty$). This effectively models the semi-infinite nature of the substrates shown in Fig. \ref{substrates}, which are blanket coated with 80 nm Au and 5 nm Ti. Numerical fits to the data are provided in the Supporting Information and match to well within 1$\%$ of the analytical fits provided in Fig. \ref{substrates}. Consequently, we consider the model used in this work to be valid for fits to the thermal properties of Si when the geometry is confined in the radial direction (e.g. Si micropillars). 

\section{Materials Synthesis and Morphology Characterization}

\begin{wrapfigure}[11]{r}{80mm}
    \centering
    \includegraphics[scale=0.8]{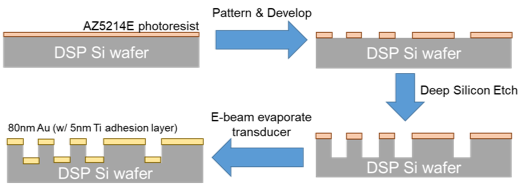}
    \caption{\small{\color{Gray}Schematic representation of the process flow to prepare the pillar arrayed investigated in this work.}}
    \label{Fab}
\end{wrapfigure}

The micro-pillars were fabricated using the following procedure (depicted schematically in Fig. \ref{Fab}). AZ5214E photoresist was spun on to the surface of a double-side polished silicon wafer (with spin rate of 4000 rpm, leading to a photoresist thickness of 1.7 $\mu$m, which was soft baked at 120$^{\circ}$C for 45 s). The patterned features were then written using a Heidelberg VPG200++ laser writer. The laser power was kept low enough for image reversal to successfully occur. The wafer was then baked at 110$^{\circ}$C for 60 s to cross-link the photoresist on the written features, and the entire wafer was flood exposed with a UV lamp with ample dosage to successfully complete the image reversal process. The wafer was then developed in 4:1 diluted AZ400K photoresist developer for 30 s. The photoresist that was not exposed during the laser write, but that was exposed during the flood exposure washed away in the developer, whereas on the written features, the photoresist remained. The wafer was then cleaned and placed in a deep silicon etch tool (PlasmaTherm DSE), where the silicon was etched away to a depth of 29.5 $\pm$ 1 $\mu$m everywhere but where the photoresist remained. Then the wafer was soaked in acetone to remove the photoresist from the written features, and the surface was cleaned in solvent rinse. Finally, a blanket-coating of 5 nm Ti and 80 nm Au was deposited by electron beam evaporation on the surface to serve as the FDTR transducer layer. 

\begin{wrapfigure}[21]{r}{80mm}
    \centering
    \includegraphics{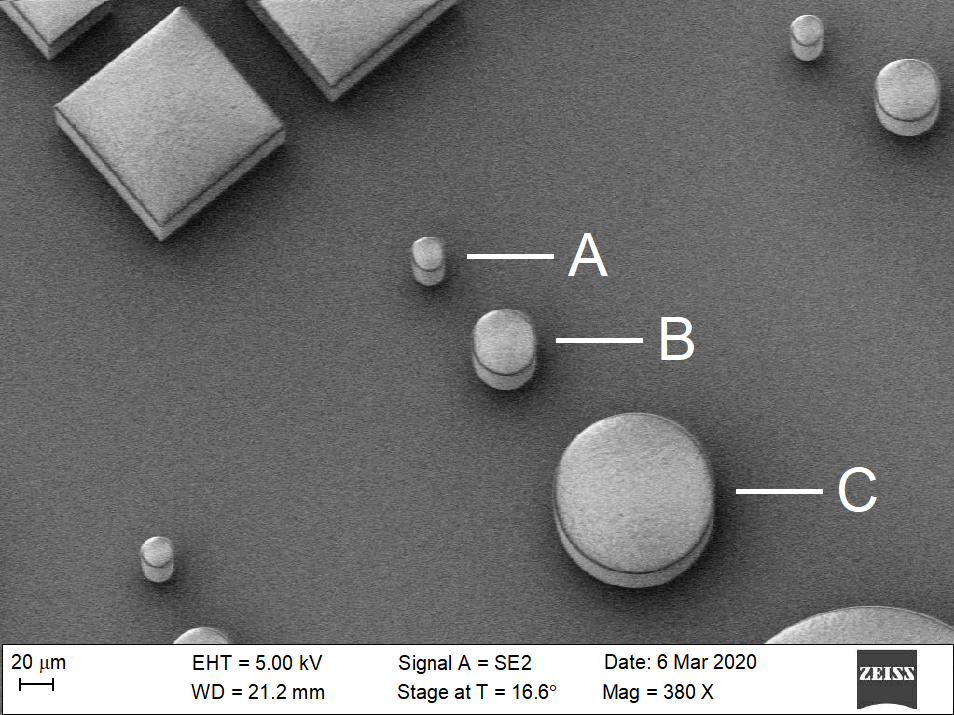}
    \caption{\small{\color{Gray}Scanning Electron Microscope image of pillar array used for this work. Note: only the three smallest circular pillars are shown, with diameters of (A) 20 $\mu$m, (B) 40 $\mu$m and (C) 100 $\mu$m.}}
    \label{SEMarray}
\end{wrapfigure}

SEM images of the micropillars used in this work are provided in Fig. \ref{SEMarray}. We note that while we only use a single pillar for characterization in this proceedings, an array of pillars was fabricated for use in future work. As shown, square pillars were also created but are not reported on in this work.

For this study, pillar C (r$_{pillar}$ = 50 $\mu$m) is used and measured with pump and probe radii of 16.8 $\mu$m and 12.8 $\mu$m, respectively. The ratio of pillar radius to pump radius (R = r$_{pillar}$/w$_{0}$) is therefore R $\approx$ 3. The pillar height is also measured by SEM as h $\approx$ 31 $\mu$m (see Supporting Information). Finally, electron beam (e-beam) evaporation is used to deposit an 80 nm Au transducer above the surface of the pillars for thermal characterization with FDTR.

In this section, we provide quantitative results that suggest the phase lag (i.e. temperature) response at the sample surface is substantially different for the confined geometry case (e.g. when r $\rightarrow$ $\infty$) across a wide range of frequencies. After we demonstrate this quantitative difference, we use our numerical fitting routine to determine the thermal conductivity of Si when in micropillar form in order to demonstrate the utility of the technique {\it and} distinguish it from the analytical solution. Consequently, we show that (1) the analytical solution is insufficient to capture the phase lag (and therefore the temperature response) at the sample surface for confined geometries and (2) a unique numerical fitting routine can be used to capture the physics that govern thermal transport in novel geometries, paving the way for the thermal characterization of micro- and nanoscale  materials with thermal properties that are expected to {\it depend on material geometry}, such as ultradrawn glass fibers and laser gain media.

\subsection{Expected Phase Lag Response for Confined Geometries}
\begin{figure*}[h!]
    \centering
    \includegraphics[scale=0.8]{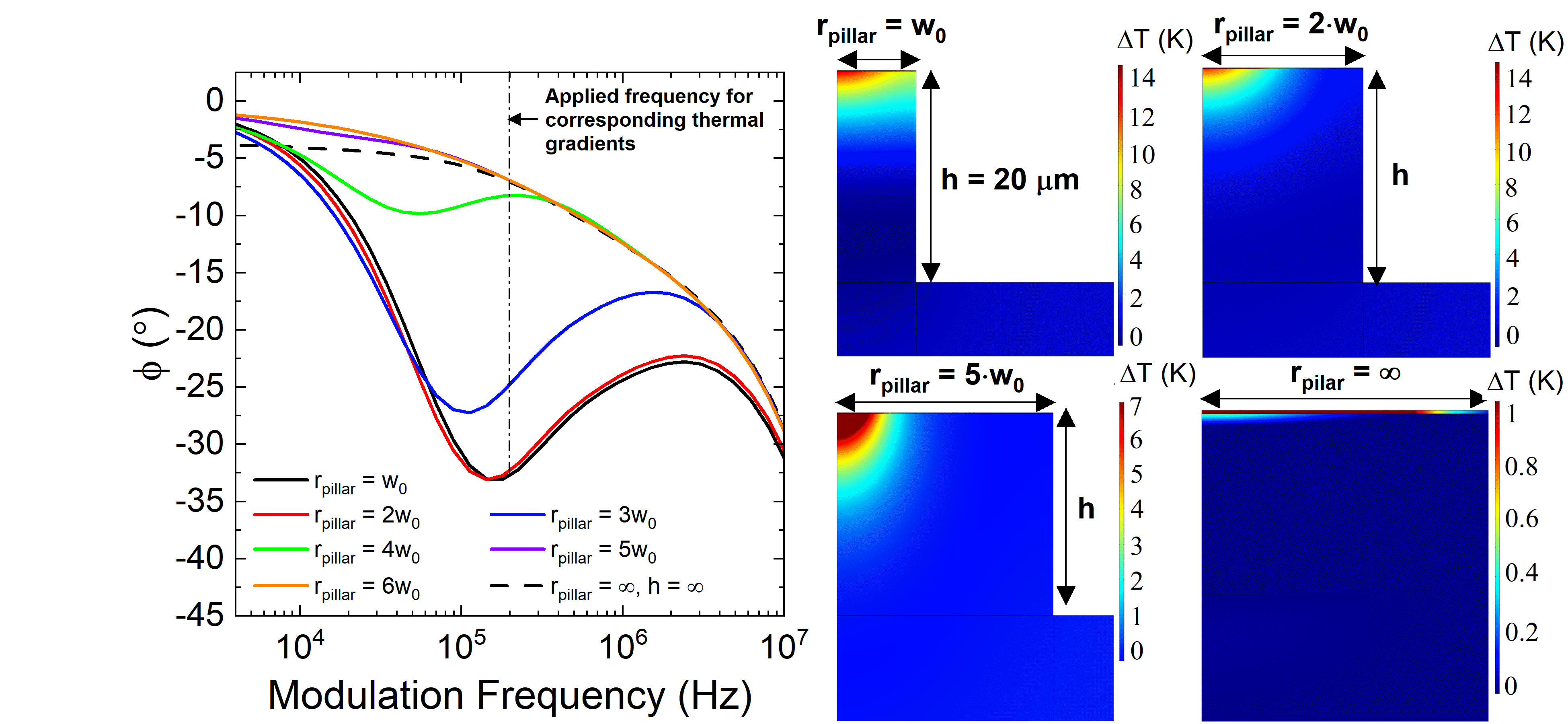}
    \caption{({\it left}) Expected phase lag at the transducer surface as a function of modulation frequency for different values of pillar radius (relative to the pump radius, w$_0$), ({\it right}) Cross-sectional view of the temperature rise within the sample at a modulation frequency of $\omega$ = 4 kHz (relative to the initial temperature with no applied power, T$_0$) with an applied power of A$_0$ = 50 mW.}
    \label{VarRad}
\end{figure*}

We computationally model a Si pillar with a 20 $\mu$m height in order to highlight the expected effects that pillar diameter has on the phase lag at the sample surface. The pump and probe diameters we use are 7.55 $\mu$m and 1.55 $\mu$m, respectively, which is consistent with those used in \cite{yang2014thermal} (and therefore provides secondary verification of the numerical results). The thermal boundary conductance is fixed at G = 35 MW m$^{-2}$ K$^{-1}$ (consistent with \cite{yang2014thermal}), $\kappa$ = 144 W m$^{-1}$ K$^{-1}$ and C$_v$ = 1.65 MJ m$^{-3}$ K$^{-1}$ (both consistent with values reported in \cite{braun2019steady} for Si). To demonstrate the power of the numerical model, we plot the phase lag as a function of modulation frequency in Fig. \ref{VarRad} as r$_{pillar}$ is varied from r$_{pillar}$ = w$_0$ to 6$\cdot$w$_0$.

As shown in Fig. \ref{VarRad}, the phase lag in the plot on the left is drastically altered as r$_{pillar}$ $\rightarrow$ w$_0$, and begins to converge on with the analytical solution for a semi-infinite Si substrate (black dashed line) at high frequencies. This makes physical sense in that, at higher frequencies, the thermal penetration depth is reduced \cite{braun2017upper} and therefore becomes less influenced by the confinement of heat at the pillar boundary. At a low modulation frequencies (e.g. $\omega$ = 4 kHz, as used for the average temperature images in the right of Fig. \ref{VarRad}), one can see that the influence of the boundary on the temperature distribution becomes more pronounced as r$_{pillar}$ $\rightarrow$ w$_0$. 

\begin{figure*}
    \centering
    \includegraphics[scale=0.85]{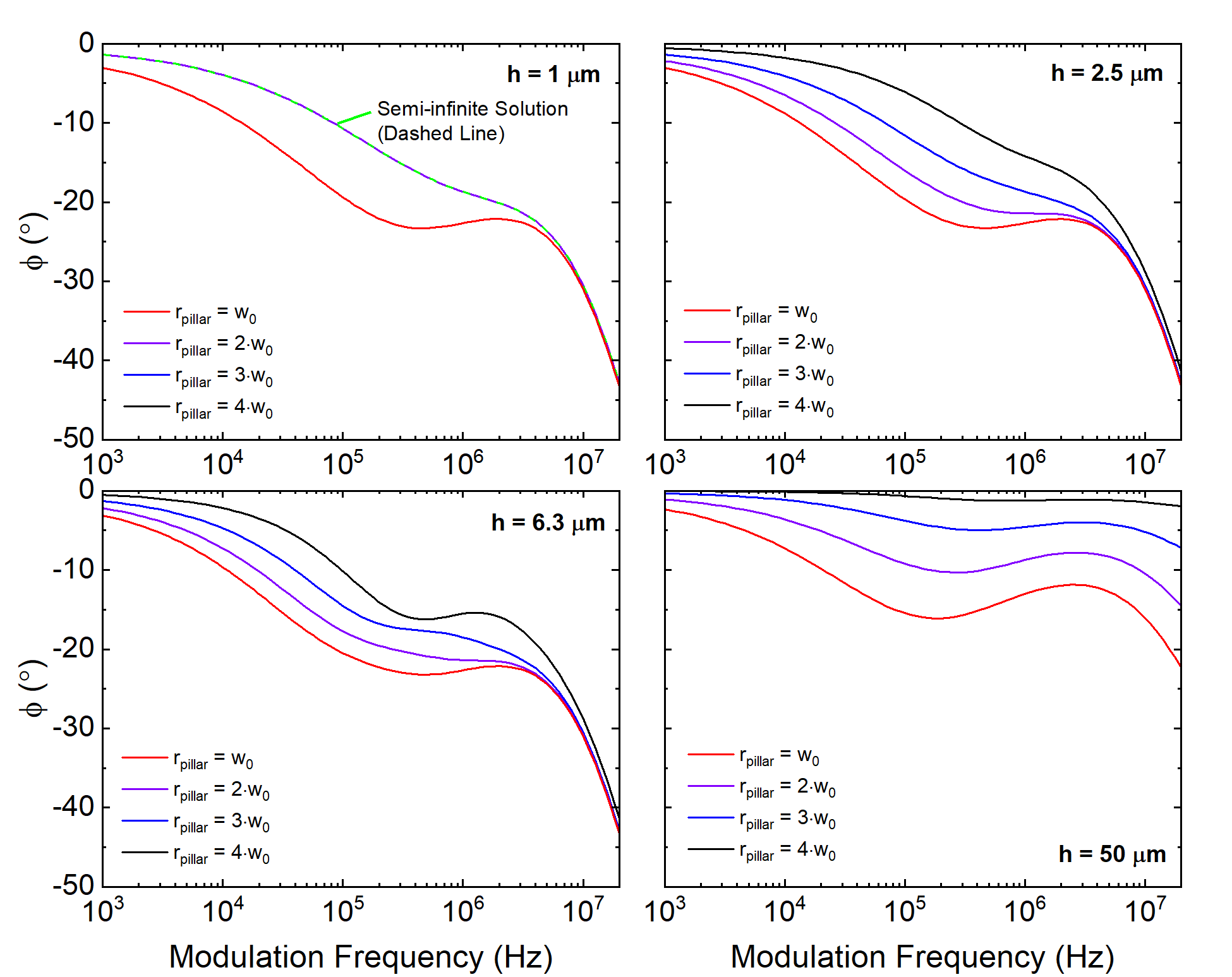}
    \caption{\small{\color{Gray}Phase lag vs. modulation frequency for ({\it top left}) h$_{pillar}$ = 1 $\mu$m, ({\it top right}) h$_{pillar}$ = 2.5 $\mu$m, ({\it bottom left}) h$_{pillar}$ = 6.3 $\mu$m and ({\it bottom right}) h$_{pillar}$ = 50 $\mu$m.}}
    \label{heights}
\end{figure*}

The height of the pillar is also expected to alter the phase lag at the sample surface relative to the case of a semi-infinite substrate. Here, the expected phase lag is simulated with heights that range from h$_{pillar}$ = 1 $\mu$m to 50 $\mu$m. Figure \ref{heights} provides the relative impact of pillar height on phase lag across a range of pillar radii (r$_{pillar}$ = w$_0$ to 4$\cdot$w$_0$).

Figure \ref{heights} reveals that for comparatively low pillar heights and large pillar radii, the phase lag at the sample surface approaches what we would expect to see for measurements on bulk substrates. This is evident in the top left plot of Fig. \ref{heights}, where pillar radii greater than 2$\cdot$w$_0$ sit neatly on the phase lag curve for semi-infinite Si (green dashed line). On the other hand, as the Si micropillar height becomes more pronounced, the phase lag at the sample surface begins to deviate drastically from the phase lag in the semi-infinite case. This is consistent with what we would expect physically; that is, as the pillar height becomes smaller, the semi-infinite substrate below it has an increasing impact on the temperature response at the sample surface. As before, a decreasing pillar radius also results in increasing deviation from the phase lag. Because these deviations are so apparent, we expect to be sensitive to the geometry of the pillar itself, permitting accurate experimental measurements of thermal transport properties via FDTR.

\subsection{Experimental Measurements of Si Micropillars with Numerical Fitting Routine}
Here we provide several measurements of the phase lag at the sample surface for Si micropillars that are between 28 and 33 $\mu$m tall and have varying radii. For this work, we utilize the one of the circular micropillars shown in Fig. \ref{SEMarray} (pillar C with r$_{pillar}$ = 49.98 $\mu$m and h$_{pillar}$ = 29 $\mu$m as measured via SEM). We note that we can vary the pump and probe radii used in our FDTR measurements by changing the objective lens magnification. This study utilizes a 20x objectives to establish pump and probe radii of 16.8 $\mu$m and 12.8 $\mu$m, respectively. 

As previously described, our numerical technique is used to extract the thermal properties of the pillar material below the transducer. The phase lag is fit to measured FDTR data with an applied pump power of A$_0$ = 7 mW, which produces a $\Delta$T$<$ 1 K at the transducer surface for our lowest modulation frequency (1 kHz). As shown, the extracted thermal conductivity of the Si micropillar measured in this work is $\kappa$ $\approx$ 145 W $\cdot$ m$^{-1}$ $\cdot$ K $^{-1}$, which remains consistent with the wider literature and within the measurement uncertainty for the conditions described in this work (U$_{\kappa}$ = 9.5$\%$) relative to the measurements made on semi-infinite substrates.

\begin{figure}[h!]
    \centering
    \includegraphics[scale=1.4]{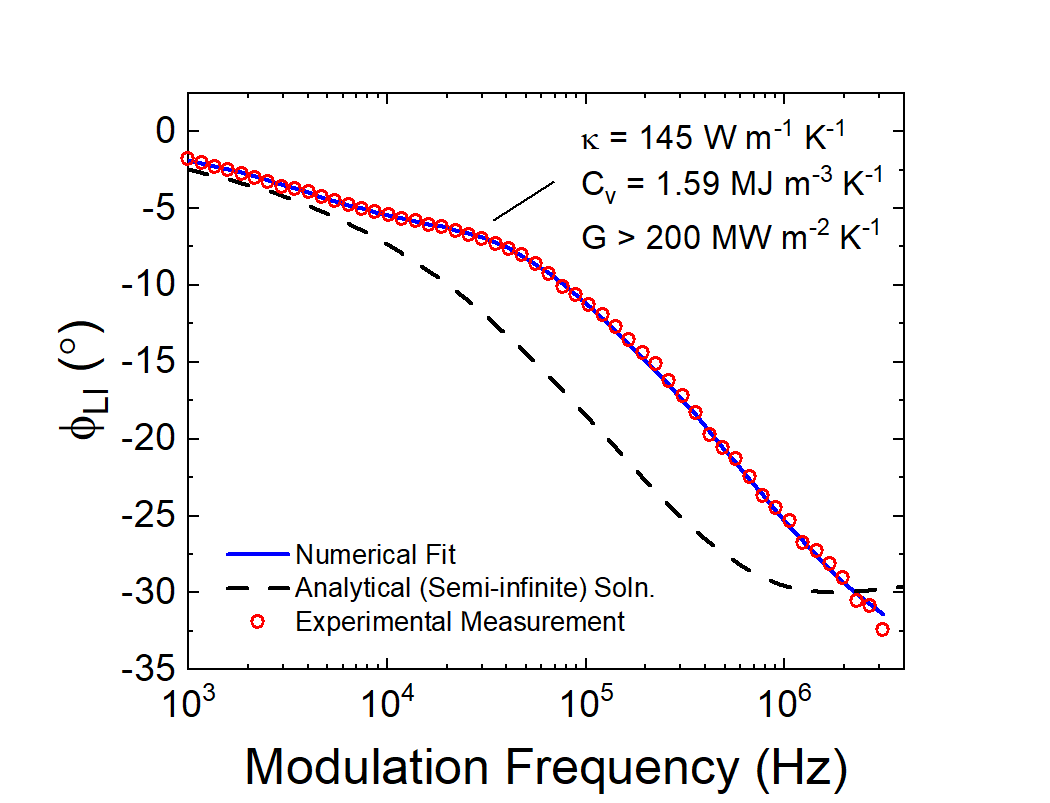}
    \caption{\small{\color{Gray}Numerical fit to the thermal properties of the Si micropillar and comparison to the expected phase lag for a completely semi-infinite geometry.}}
    \label{fit}
\end{figure}

We note here that an attempted fit of the data using the analytical model was not possible (shown in Fig. \ref{fit} is the {\it expected} phase lag when the sample is semi-infinite in order to clearly identify that the micropillar geometry has an impact on the phase lag at the transducer surface). A comparison between the obtained thermal properties for the semi-infinite geometry and the micropillar geometry are provided in Table \ref{tab}.

\begin{table}[h!]
    \centering
    \resizebox{\columnwidth}{!}{%
    \begin{tabular}{l|ccc}
        \hline
        \hline
         {\bf Property} & {\bf SI Analytical} & {\bf SI Numerical} & {\bf MP Numerical} \\
         \hline
         $\kappa$ (W m$^{-1}$ K$^{-1}$) & 139 $\pm$ 7 & 139.2 $\pm$ 7.1 & 145 $\pm$ 10\\ 
         C$_v$ (MJ m$^{-3}$ K$^{-1}$) & 1.5 $\pm$ 0.3 & 1.6 $\pm$ 0.2 & 1.59 $\pm$ 0.2\\
         \hline
         \hline
    \end{tabular}
    }
    \caption{Comparison of thermal properties for semi-infinite (SI) samples using analytical and numerical fitting routines to micropillar (MP) numerical fitting routine. G is not shown due to lack of sufficient sensitivity.}
    \label{tab}
\end{table}

Table \ref{tab} suggests that the formulation described here appropriately captures the thermal properties of materials whose geometries are confined, and are quite different from those solutions which use the standard analytical fitting routine. Consequently, current analytical formulations \cite{schmidt2009frequency} for resolving the thermal conductivity of materials having non semi-infinite geometries are insufficient to resolve their thermal properties. 

To provide further evidence that the fitting routine is able to capture the effects of the adiabatic boundary on the phase response at the upper surface of the pillar, we use our fitting routine to determine the thermal conductivity of Si micropillars having different diameters. 

\begin{figure}[h!]
    \centering
    \includegraphics[scale=0.45]{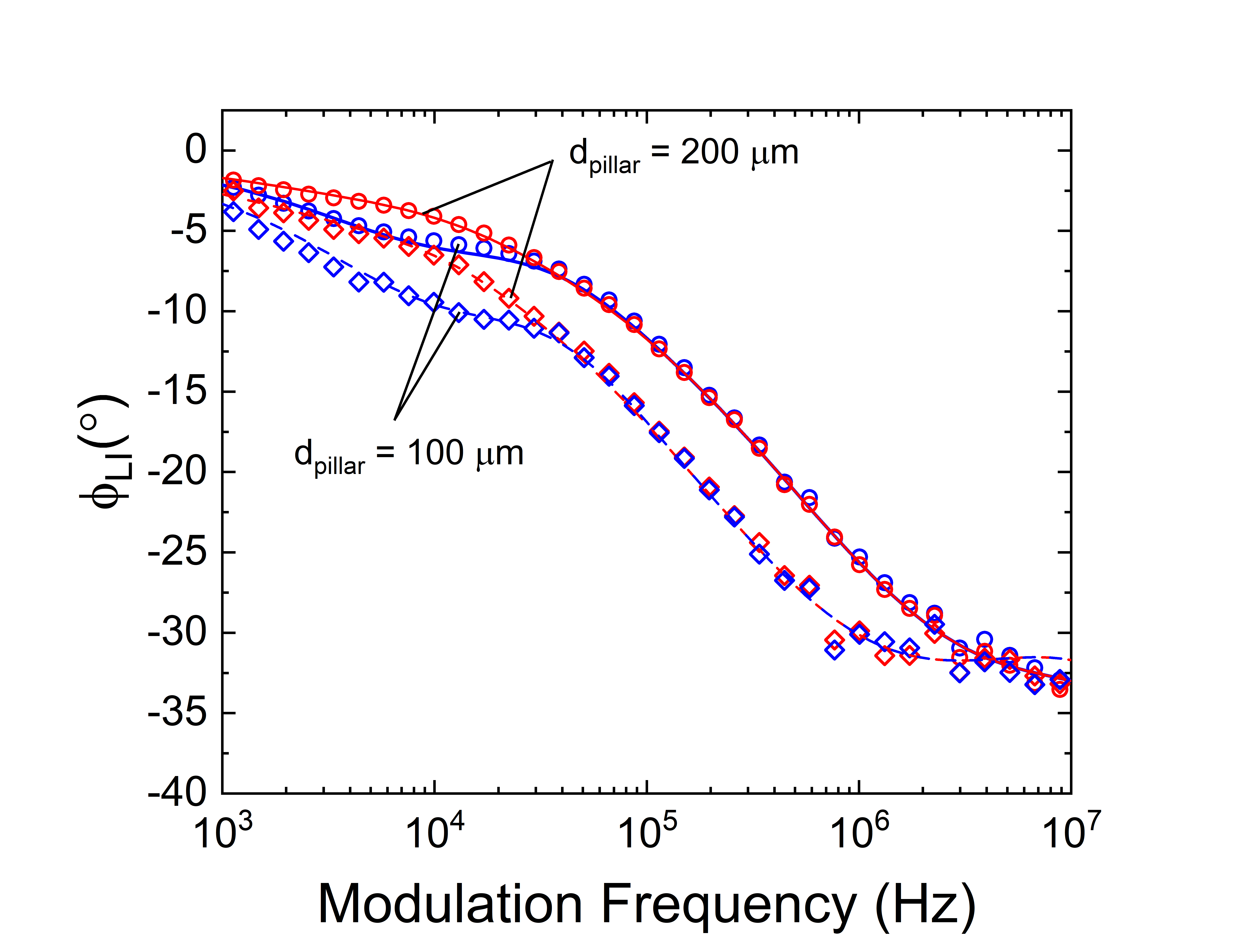}
    \caption{\small{\color{Gray}Phase lag vs. frequency for micropillars with d$_{pillar}$ = 100 $\mu$m (blue) and 200 $\mu$m (red) and pump beam diameters of w$_0$ = 21.2 $\mu$m (circles) and 35.7 $\mu$m (diamonds). Solid and dashed lines represent the numerical fits to our data for the 21.2 $\mu$m and 35.7 $\mu$m diameter pump beams, respectively.}}
    \label{morefits}
\end{figure}

\begin{figure}[h!]
    \centering
    \includegraphics[scale=0.45]{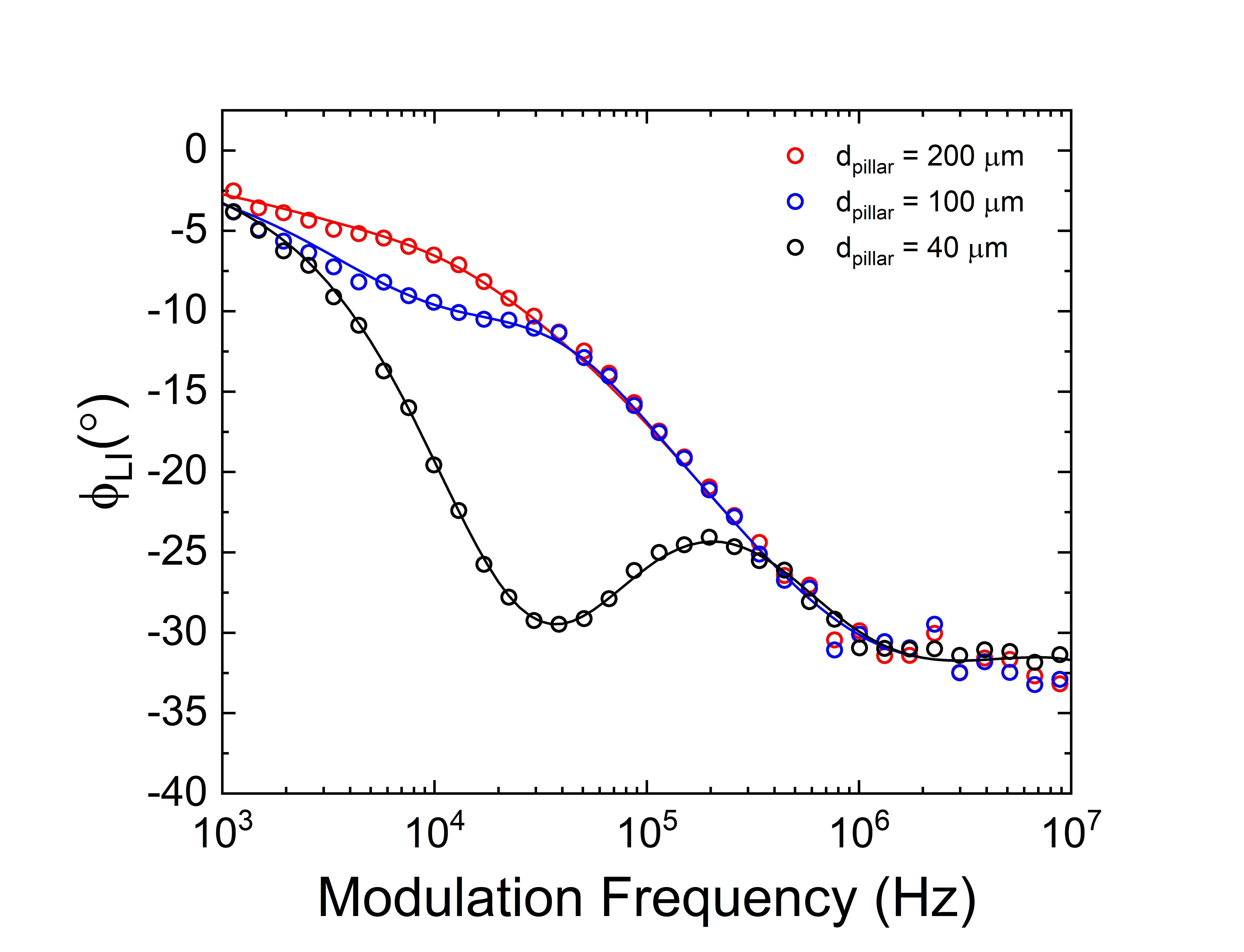}
    \caption{\small{\color{Gray}Phase lag vs. frequency for micropillars of varying size with w$_0$ = 35.7 $\mu$m.}}
    \label{varies}
\end{figure}

\noindent We also utilize different pump sizes than those used in Fig. \ref{fit} (in this case, w$_0$ = 21.2 $\mu$m, and w$_0$ = 35.7 $\mu$m). We first compare the impact of the relative size of the pump beams on the phase lag for micropillar diameters of 100 $\mu$mand 200 $\mu$m) in order to highlight the sensitivity of the phase lag to the presence of confined boundaries, as shown in Fig. \ref{morefits}.

In Fig. \ref{morefits}, the discrepancy between the phase lag for each of the two pump diameters is more pronounced for the pump beam with the larger diameter. This suggests that as the size of the pump beam is increased relative to the size of the pillar, the boundaries start to play a larger role on the temperature at the surface of the transducer.

We observe the same phenomenon when we hold the pump diameter constant and reduce the pillar cross-section, as shown in Fig. \ref{varies}. In Fig. \ref{varies}, the phase lag is shown to be dependent on the ratio of the pump diameter to the diameter of the pillar. As the pump beam becomes more confined by the geometry of the pillar, its impact on the phase lag in the low frequency regime is more pronounced. We note that all fits resulted in a thermal conductivity of $\kappa$ = 145 W/m$\cdot$K, which is consistent with that reported in the wider literature. This fitting routine is therefore critical when characterizing materials that are expected to have thermal properties that are dependent on finite geometries, such as ultra-drawn glass and polymer fibers \cite{shen2010polyethylene,matsuo1988dynamic} and laser gain media.

\section*{CONCLUSIONS}
In this work, we develop a numerical model that is capable of fitting the phase lag of FDTR measurements on samples having non semi-infinite geometries. We show that, for a known geometry, we can extract accurate values of thermal conductivity, volumetric heat capacity and the thermal boundary conductance across the transducer/material interface for a non-typical geometry (in this case, a Si micropillar). The impact of the substrate is included to incorporate the impact of thermal spreading at low pump beam modulation frequencies, which we clearly demonstrate in a numerical parametric analysis. Given this new fitting paradigm, we expect thermal scientists to be able to characterize materials having geometry-dependent thermal properties, to include ultra-drawn glass fibers and laser gain media.\\

\section*{ACKNOWLEDGEMENTS}
RJW and BFD would like to acknowledge primary financial support from Mr. Peter Morrison and the Office of Naval Research (N001420WX00381). We  also thank Dr. Mark Spector and ONR for equipment support (N0001420WX01170). 

\nolinenumbers

\bibliography{library}

\begin{thebibliography}{10}

\bibitem{jiang2018tutorial}
Puqing Jiang, Xin Qian, and Ronggui Yang.
\newblock Tutorial: Time-domain thermoreflectance (tdtr) for thermal property
  characterization of bulk and thin film materials.
\newblock {\em Journal of Applied Physics}, 124(16):161103, 2018.

\bibitem{yan2014thermal}
Rusen Yan, Jeffrey~R Simpson, Simone Bertolazzi, Jacopo Brivio, Michael Watson,
  Xufei Wu, Andras Kis, Tengfei Luo, Angela~R Hight~Walker, and Huili~Grace
  Xing.
\newblock Thermal conductivity of monolayer molybdenum disulfide obtained from
  temperature-dependent raman spectroscopy.
\newblock {\em ACS Nano}, 8(1):986--993, 2014.

\bibitem{tong2006reexamining}
Tao Tong and Arun Majumdar.
\newblock Reexamining the 3-omega technique for thin film thermal
  characterization.
\newblock {\em Review of Scientific Instruments}, 77(10):104902, 2006.

\bibitem{kang2008two}
Kwangu Kang, Yee~Kan Koh, Catalin Chiritescu, Xuan Zheng, and David~G Cahill.
\newblock Two-tint pump-probe measurements using a femtosecond laser oscillator
  and sharp-edged optical filters.
\newblock {\em Review of Scientific Instruments}, 79(11):114901, 2008.

\bibitem{feser2014pump}
Joseph~P Feser, Jun Liu, and David~G Cahill.
\newblock Pump-probe measurements of the thermal conductivity tensor for
  materials lacking in-plane symmetry.
\newblock {\em Review of Scientific Instruments}, 85(10):104903, 2014.

\bibitem{ge2006thermal}
Zhenbin Ge, David~G Cahill, and Paul~V Braun.
\newblock Thermal conductance of hydrophilic and hydrophobic interfaces.
\newblock {\em Physical Review Letters}, 96(18):186101, 2006.

\bibitem{lyeo2006thermal}
Ho-Ki Lyeo and David~G Cahill.
\newblock Thermal conductance of interfaces between highly dissimilar
  materials.
\newblock {\em Physical Review B}, 73(14):144301, 2006.

\bibitem{losego2012effects}
Mark~D Losego, Martha~E Grady, Nancy~R Sottos, David~G Cahill, and Paul~V
  Braun.
\newblock Effects of chemical bonding on heat transport across interfaces.
\newblock {\em Nature Materials}, 11(6):502--506, 2012.

\bibitem{cahill2004analysis}
David~G Cahill.
\newblock Analysis of heat flow in layered structures for time-domain
  thermoreflectance.
\newblock {\em Review of Scientific Instruments}, 75(12):5119--5122, 2004.

\bibitem{zhu2010ultrafast}
Jie Zhu, Dawei Tang, Wei Wang, Jun Liu, Kristopher~W Holub, and Ronggui Yang.
\newblock Ultrafast thermoreflectance techniques for measuring thermal
  conductivity and interface thermal conductance of thin films.
\newblock {\em Journal of Applied Physics}, 108(9):094315, 2010.

\bibitem{schmidt2009frequency}
Aaron~J Schmidt, Ramez Cheaito, and Matteo Chiesa.
\newblock A frequency-domain thermoreflectance method for the characterization
  of thermal properties.
\newblock {\em Review of Scientific Instruments}, 80(9):094901, 2009.

\bibitem{favaloro2015characterization}
T~Favaloro, J-H Bahk, and A~Shakouri.
\newblock Characterization of the temperature dependence of the
  thermoreflectance coefficient for conductive thin films.
\newblock {\em Review of Scientific Instruments}, 86(2):024903, 2015.

\bibitem{wilson2012thermoreflectance}
RB~Wilson, Brent~A Apgar, Lane~W Martin, and David~G Cahill.
\newblock Thermoreflectance of metal transducers for optical pump-probe studies
  of thermal properties.
\newblock {\em Optics Express}, 20(27):28829--28838, 2012.

\bibitem{braun2019steady}
Jeffrey~L Braun, David~H Olson, John~T Gaskins, and Patrick~E Hopkins.
\newblock A steady-state thermoreflectance method to measure thermal
  conductivity.
\newblock {\em Review of Scientific Instruments}, 90(2):024905, 2019.

\bibitem{kimling2017thermal}
Judith Kimling, Andr{\'e} Philippi-Kobs, Jonathan Jacobsohn, Hans~Peter Oepen,
  and David~G Cahill.
\newblock Thermal conductance of interfaces with amorphous sio2 measured by
  time-resolved magneto-optic kerr-effect thermometry.
\newblock {\em Physical Review B}, 95(18):184305, 2017.

\bibitem{chen2017effects}
Xi~Chen, Karalee Jarvis, Sean Sullivan, Yutao Li, Jianshi Zhou, and Li~Shi.
\newblock Effects of grain boundaries and defects on anisotropic magnon
  transport in textured sr$_14$ cu$_24$ o$_41$.
\newblock {\em Physical Review B}, 95(14):144310, 2017.

\bibitem{szwejkowski2017molecular}
Chester~J Szwejkowski, Ashutosh Giri, Ronald Warzoha, Brian~F Donovan, Bryan
  Kaehr, and Patrick~E Hopkins.
\newblock Molecular tuning of the vibrational thermal transport mechanisms in
  fullerene derivative solutions.
\newblock {\em ACS Nano}, 11(2):1389--1396, 2017.

\bibitem{johnson2013direct}
Jeremy~A Johnson, AA~Maznev, John Cuffe, Jeffrey~K Eliason, Austin~J Minnich,
  Timothy Kehoe, Clivia M~Sotomayor Torres, Gang Chen, and Keith~A Nelson.
\newblock Direct measurement of room-temperature nondiffusive thermal transport
  over micron distances in a silicon membrane.
\newblock {\em Physical Review Letters}, 110(2):025901, 2013.

\bibitem{kim2017elastic}
Taeyong Kim, Ding Ding, Jong-Hyuk Yim, Young-Dahl Jho, and Austin~J Minnich.
\newblock Elastic and thermal properties of free-standing molybdenum disulfide
  membranes measured using ultrafast transient grating spectroscopy.
\newblock {\em APL Materials}, 5(8):086105, 2017.

\bibitem{minnich2012determining}
Austin~J Minnich.
\newblock Determining phonon mean free paths from observations of
  quasiballistic thermal transport.
\newblock {\em Physical Review Letters}, 109(20):205901, 2012.

\bibitem{warzoha2019nanoscale}
Ronald~J Warzoha, Brian~F Donovan, Nicholas~T Vu, James~G Champlain, Shawn
  Mack, and Laura~B Ruppalt.
\newblock Nanoscale thermal transport in amorphous and crystalline gete
  thin-films.
\newblock {\em Applied Physics Letters}, 115(2):023104, 2019.

\bibitem{donovan2019elimination}
Brian~F Donovan, Ronald~J Warzoha, R~Bharath Venkatesh, Nicholas~T Vu, Jay
  Wallen, and Daeyeon Lee.
\newblock Elimination of extreme boundary scattering via polymer thermal
  bridging in silica nanoparticle packings: Implications for thermal
  management.
\newblock {\em ACS Applied Nano Materials}, 2(10):6662--6669, 2019.

\bibitem{wang2011thermal}
Zhaojie Wang, Joseph~E Alaniz, Wanyoung Jang, Javier~E Garay, and Chris Dames.
\newblock Thermal conductivity of nanocrystalline silicon: importance of grain
  size and frequency-dependent mean free paths.
\newblock {\em Nano Letters}, 11(6):2206--2213, 2011.

\bibitem{yang2014thermal}
Jia Yang, Elbara Ziade, Carlo Maragliano, Robert Crowder, Xuanye Wang, Marco
  Stefancich, Matteo Chiesa, Anna~K Swan, and Aaron~J Schmidt.
\newblock Thermal conductance imaging of graphene contacts.
\newblock {\em Journal of Applied Physics}, 116(2):023515, 2014.

\bibitem{braun2017upper}
Jeffrey~L Braun and Patrick~E Hopkins.
\newblock Upper limit to the thermal penetration depth during modulated heating
  of multilayer thin films with pulsed and continuous wave lasers: A numerical
  study.
\newblock {\em Journal of Applied Physics}, 121(17):175107, 2017.

\bibitem{shen2010polyethylene}
Sheng Shen, Asegun Henry, Jonathan Tong, Ruiting Zheng, and Gang Chen.
\newblock Polyethylene nanofibres with very high thermal conductivities.
\newblock {\em Nature Nanotechnology}, 5(4):251, 2010.

\bibitem{matsuo1988dynamic}
Masaru Matsuo, Chie Sawatari, and Tomoko Ohhata.
\newblock Dynamic mechanical studies on the crystal dispersion using ultradrawn
  polyethylene films.
\newblock {\em Macromolecules}, 21(5):1317--1324, 1988.

\end{thebibliography}

\bibliographystyle{unsrt}

\end{document}